\documentstyle[12pt]{article}

\def\overlay#1#2{\ifmmode%
\setbox0=\hbox{$#1$}%
\setbox1=\hbox to\wd0{\hss$#2$\hss}\else%
\setbox0=\hbox{#1}%
\setbox1=\hbox to\wd0{\hss#2\hss}\fi%
 #1\hskip-\wd0\box1 }

\begin{document}
\hfill\vbox{
\hbox{NUHEP-TH-94-24}
\hbox{hep-ph/9409434}
\hbox{September 1994} }\par
\thispagestyle{empty}

\begin{center}
{\Large \bf Solution to the Perturbative Infrared Catastrophe\\
of Hot Gauge Theories}

\vspace{0.15in}

Eric Braaten\\

{\it Department of Physics and Astronomy, Northwestern University,
Evanston, IL 60208}
\end{center}

\begin{abstract}
The free energy of a nonabelian gauge theory at high temperature $T$
can be calculated to order $g^5$ using resummed perturbation theory,
but the method breaks down at order $g^6$.
A new method is developed for calculating the free energy to arbitrarily
high accuracy in the high temperature limit.  It involves
the construction of a sequence of two effective field theories
by first integrating out the momentum scale $T$ and then integrating out the
momentum scale $g T$.  The free energy decomposes into the sum of
three terms, corresponding to the momentum scales $T$, $gT$, and $g^2T$.
The first term can be calculated as
a perturbation series in $g^2(T)$, where $g(T)$ is the running coupling
constant.  The second term in the free energy can be calculated as a
perturbation series in $g(T)$, beginning at order $g^3$.
The third term can also be expressed as a series in $g(T)$
beginning at order $g^6$, with coefficients that can be
calculated using lattice simulations of 3-dimensional QCD.
Leading logarithms of $T/(gT)$ and of $gT/(g^2T)$ can be summed up using
renormalization group equations.

\end{abstract}
\newpage

The perturbative infrared catastrophe of a nonabelian gauge theory
at high temperature  is one of the most important unsolved
problems in thermal field theory.  The problem, which was first identified
by Linde in 1979 \cite{linde,gpy}, is that perturbative calculations of the
free energy of a nonabelian gauge theory at high temperature $T$
with weak coupling $g$ break down at order $g^6$.
In contrast, there seems to be no obstacle to
calculating the free energy of an abelian gauge theory like QED to arbitrarily
high order in the coupling constant $e$.  This problem casts a shadow
over all applications of QCD perturbation theory to describe a quark-gluon
plasma at high temperature, because the catastrophe afflicts any observable
at sufficiently high order in $g$.  The catastrophe also arises in
applications of the electroweak gauge theory and grand-unified theories
to describe phase transitions in the early universe, since
they also involve nonabelian gauge theories at high temperature.
The catastrophe can be avoided for static observables
in hot QCD by calculating them directly using lattice simulations,
but the computational resources that are required increase rapidly
with $T$.  Moreover, there are
many problems, such as the calculation of dynamical observables or
the calculation of static observables at nonzero baryon density,
which are more easily addressed by perturbative methods than by
lattice simulations.  For these reasons, it is important to solve the
problem of the perturbative infrared catastrophe.

In this Letter, we solve the problem by constructing a sequence of two
effective field theories.  The first effective theory,
which is obtained by integrating out the momentum scale $T$,
is equivalent to thermal QCD at length scales of order $1/(gT)$
or larger.  The second effective theory, which is obtained by integrating
out the scale $gT$ from the first effective theory,
is equivalent to thermal QCD at length scales
of order $1/(g^2T)$ or larger.   If $T$ is sufficiently large,
the parameters of the two effective theories can be calculated as
perturbation series in the running coupling constant $g(T)$.
The second effective theory is inherently nonperturbative, so that the
effects from the scale $g^2T$ must be calculated by lattice simulations.

To explain the origin of the infrared catastrophe, we first consider
the case of the abelian gauge theory QED.  The leading term in the free
energy is that of an ideal gas of electrons, positrons, and photons.
The first correction due to the electromagnetic interaction is of order
$e^2$, and can be calculated using ordinary perturbation theory.
Beyond that order, infrared divergences arise in perturbation theory
and it is necessary to sum infinite classes of diagrams.
The order $e^3$ correction to the free energy can be calculated by resumming
``ring diagrams'', as in the classic calculation by Gell-Mann
and Brueckner for a degenerate electron gas \cite{gellmann}.
This method was used to calculate the order $e^4$ correction to the
free energy of a degenerate electron gas in the relativistic limit
\cite{mclerran}.  An equivalent resummation method \cite{parwani1} was used to
calculate the free energy to order $g^4$ for a massless scalar field with
a $\phi^4$ interaction in the high temperature limit \cite{fst}.
More recently, this
method was used to calculate the corrections to the free energy for hot QED
of orders $e^4$ \cite{parwani-coriano} and $e^5$ \cite{parwani2}.
The only apparent obstacle to calculating to
still higher order in $e$ is the increasing complexity of the
sums and integrals that are encountered.

We now consider the case of hot QCD.
The resummation of ring diagrams allows the free energy to be
calculated to order $g^5$.  The free energy for pure-glue QCD
was recently calculated to order $g^4$ by Arnold and Zhai \cite{arnold-zhai}
using an equivalent resummation method \cite{arnold-espinoza}.
However, at order $g^6$, there are contributions
to the free energy that can only be calculated using nonperturbative methods.
To see where they arise, recall that static quantities
in a thermal field theory can be calculated using the imaginary-time
formalism, in which the propagator for a massless field is
$1/(\omega_n^2 + {\bf k}^2)$, where $\omega_n$
takes on the discrete values $\omega_n = 2 n \pi T$ for bosons
and $\omega_n = (2n+1)\pi T$ for fermions.  The only modes that can
propagate over distances much larger than $1/T$ are the $n = 0$ modes
of the bosons.  These are therefore the only modes that can give
rise to infrared divergences that cause a breakdown of perturbation theory.
With the resummation of ring diagrams,  the chromoelectric
zero modes (``electrostatic gluons'') acquire a mass of order $gT$,
while chromomagnetic zero modes (``magnetostatic gluons'') remain massless.
When restricted to the chromomagnetic zero modes, the action of QCD
reduces to that of 3-dimensional Euclidean QCD with coupling constant $g^2T$.
The perturbation expansion for this theory is hopelessly afflicted with
infrared divergences, but it is well-behaved
nonperturbatively with a mass gap of order $g^2T$.  Using simple
dimensional analysis, one can show that the contribution
to the free energy density from these modes is $(g^2 T)^3 T$ multiplied by a
coefficient that can only be obtained by a nonperturbative calculation.
There is no analogous contribution in QED, because the restriction to
magnetostatic zero modes gives a free field theory.

We present the solution to the perturbative infrared catastrophe
in the context of thermal QCD, although the solution applies equally well
to any nonabelian gauge theory.
Our starting point is the imaginary-time formalism for thermal QCD,
in which static
quantities are calculated in 4-dimensional Euclidean QCD, with the
periodic Euclidean time $\tau$ having
period $\beta = 1/T$.   The partition function is
\begin{equation}
Z_{\rm QCD}(T) \;=\; \int {\cal D} A_\mu^a
\exp \left( - \int_0^\beta d \tau \int d^3x
	{1 \over 4} G^a_{\mu \nu} G^a_{\mu \nu}\right),
\label{ZQCD}
\end{equation}
where
$G^a_{\mu \nu}$ is the nonabelian field strength constructed out of the
gauge field $A^a_\mu$ with coupling constant $g$.  The free energy density
is $F = - T \log Z/V$, where $V$ is the volume of space.
It can be calculated at any temperature $T$
using lattice simulations of the functional integral.
If $T$ is large enough that the running coupling constant $g(T)$ is small,
one might expect to be able to calculate $F(T)$ as a power series in
$g^2(T)$ using ordinary perturbation theory.
However this perturbation expansion
breaks down due to infrared divergences associated with the exchange of
static gluons.  The divergences due to electrostatic gluons
can be eliminated by the resummation of ring diagrams,
but a more elegant solution is to construct
an effective field theory which reproduces the static correlation functions
of thermal QCD at distances of order $1/(gT)$ or larger.  This theory,
which we call electrostatic QCD (EQCD), contains an electrostatic
gauge field $A_0^a({\bf x})$ and a magnetostatic gauge field $A_i^a({\bf x})$.
Up to normalizations, they can be identified
with the zero-frequency modes of the gluon field $A^a_\mu({\bf x}, \tau)$
for thermal QCD in a static gauge \cite{nadkarni}.
The lagrangian for this effective field theory is
\begin{equation}
{\cal L}_{\rm EQCD} \;=\;
{1 \over 4} G^a_{ij} G^a_{ij} \;+\; {1 \over 2} (D_i A_0)^a (D_i A_0)^a
\;+\; {1 \over 2} m_{\rm el}^2 A_0^a A_0^a
	\;+\; \delta {\cal L}_{\rm EQCD},
\label{EQCD}
\end{equation}
where $G_{ij}$ is the magnetostatic field strength with coupling constant
$g_3$.   The lagrangian (\ref{EQCD}) has an $SU(3)$ gauge symmetry.
If the fields $A_0$ and $A_i$ are assigned the scaling
dimension $1/2$, then the operators shown explicitly in (\ref{EQCD})
have dimensions 3, 3, and 1.   The term $\delta {\cal L}_{\rm EQCD}$
in (\ref{EQCD}) contains all possible local gauge-invariant operators
of dimension 2 and higher that can be constructed out of $A_0$ and $A_i$.

Renormalization theory tells us that static correlation functions at distances
$R \gg 1/T$ in the full theory can be reproduced in the
effective theory with any accuracy desired
by tuning the coupling constant $g_3$, the mass parameter $m_{\rm el}^2$,
and the parameters in $\delta {\cal L}_{\rm EQCD}$ as functions of
$T$ and the ultraviolet cutoff $\Lambda$ of the effective theory \cite{lepage}.
The $\Lambda$-dependence of the parameters is cancelled by the
$\Lambda$-dependence of the loop integrals in the effective theory.
The coefficients of the operators in the effective lagrangian
(\ref{EQCD}) for EQCD can be computed
as perturbation series in $g^2(T)$.  At leading order in $g$, we have
$g_3 = g \sqrt{T}$ and $m_{\rm el}^2 = N_c g^2 T^2/3$.
The coefficients of some of the higher-dimension operators were recently
computed to leading order by Chapman \cite{chapman}.
Beyond leading order, it is particularly convenient to use a scale-invariant
regularization scheme, such as dimensional regularization, which automatically
eliminates all power ultraviolet divergences.  In this case, the
coefficient of an operator of dimension $d$ is $T^{3-d}$
multiplied by a power series in $g^2(T)$,
with coefficients that are polynomials in $\log(T/\Lambda)$.
The parameters satisfy simple renormalization group equations that can be used
to sum up leading logarithms of $T/\Lambda$.

The effective field theory EQCD can be used to calculate
the partition function of thermal QCD, as well as static correlation
functions at distance scales of order $1/(gT)$ or larger.
The partition function is
\begin{equation}
Z_{\rm QCD}(T) \;=\;
e^{- f_{\rm QCD}(\Lambda) \, T^3 \, V}
\int^{(\Lambda)} {\cal D} A_0^a {\cal D} A_i^a
\exp \left( - \int d^3x {\cal L}_{\rm EQCD} \right).
\label{ZEQCD}
\end{equation}
In the exponential prefactor, $f_{\rm QCD}T^3$  can be interpreted
as the coefficient of the unit operator which was omitted from
the effective lagrangian (\ref{EQCD}).  It depends on the ultraviolet cutoff
$\Lambda$ of the effective theory in such a way as to cancel the
cutoff dependence of the loop integrals in EQCD.

If we use dimensional regularization in $3-2\epsilon$ spacial dimensions
to cut off both ultraviolet and infrared divergences, then $f_{\rm QCD}$
is given by the ordinary diagrammatic expansion for the free energy of
thermal QCD.  It can be calculated to
next-to-next-to-leading order in $g^2$ by calculating 1-loop,
2-loop, and 3-loop diagrams without any resummation.  These
diagrams can be obtained from the results given in Ref. \cite{arnold-zhai}
by subtracting the effects of resummation.  Taking the $\overline{MS}$
scale of dimensional regularization to be $\Lambda$ and renormalizing
the coupling constant at the scale $4 \pi T$, we obtain
\begin{eqnarray}
f_{\rm QCD} &=& {(N_c^2-1) \pi^2 \over 9}
\Bigg\{ -{1 \over 5} \;+\; {N_c g^2(4 \pi T) \over 16 \pi^2}
\nonumber \\
&+& \left( {N_c g^2 \over 16 \pi^2} \right)^2
\Bigg[ - {12 \over \epsilon} - 72 \log {\Lambda \over 4 \pi T}
- 4 \gamma - {116 \over 5} - {220 \over 3} {\zeta'(-1) \over \zeta(-1)}
+ {38 \over 3} {\zeta'(-3) \over \zeta(-3)} \Bigg] \Bigg\} ,
\label{fQCD}
\end{eqnarray}
where $\gamma$ is Euler's constant and $\zeta(z)$ is the Riemann zeta
function.  The pole in $\epsilon$ arises from an infrared divergence.
The mass parameter $m_{\rm el}^2$ in (\ref{EQCD}) is obtained
to leading order in $g^2$ by matching the 1-loop propagator corrections
for electrostatic gluons in thermal QCD to the propagator correction
in EQCD given by the term $m_{\rm el}^2 A^a_0 A^a_0/2$
in the lagrangian (\ref{EQCD}):
\begin{equation}
m_{\rm el}^2 \;=\; {N_c g^2 T^2 \over 3}
\left[ 1 + \epsilon \left( 2 \log {\Lambda \over 4 \pi T} +
2 {\zeta'(-1) \over \zeta(-1)} \right) \right] .
\label{mel}
\end{equation}
Since logarithmic ultraviolet divergences in the effective theory
show up as poles in $\epsilon$, we must keep the terms of order $\epsilon$
in (\ref{mel}).

Once the parameters in ${\cal L}_{\rm EQCD}$ have been determined
to sufficiently high accuracy, the free energy density $F(T)$
can be calculated using lattice simulations of the functional integral
in (\ref{ZEQCD}).  If the coupling constant is sufficiently small,
one might also expect to be able to calculate $F(T)$ using perturbation
theory.  The relevant expansion
parameter is $g_3^2/m_{\rm el}$, which is of order $g$, so the
perturbation series is an expansion in powers of $g(T)$, rather than $g^2(T)$.
Unfortunately, the
perturbation expansion of this effective field theory breaks down due
to infrared divergences associated with magnetostatic gluons.
It is convenient to construct a second effective field
theory containing only the magnetostatic gauge field $A^a_i({\bf x})$,
which we call magnetostatic QCD (MQCD).
The lagrangian for this effective field theory is
\begin{equation}
{\cal L}_{\rm MQCD} \;=\;
{1 \over 4} G^a_{ij} G^a_{ij}
\;+\; \delta {\cal L}_{\rm MQCD} ,
\label{MQCD}
\end{equation}
where $G_{ij}$ is the magnetostatic field strength with coupling constant
$g_3'$, which differs from $g_3$ by perturbative corrections.  The term
$\delta {\cal L}_{\rm MQCD}$ includes all possible local
gauge-invariant operators of dimension 5 and higher
that can be constructed out of $A_i^a$.
Renormalization theory tells us that magnetostatic correlation functions
at distances $R \gg 1/(gT)$ or larger in EQCD can be reproduced in MQCD
to any desired accuracy by tuning the coupling constant $g_3'$
and the parameters in $\delta {\cal L}_{\rm MQCD}$ as functions of
the parameters of EQCD and the ultraviolet cutoff $\Lambda'$
of MQCD \cite{lepage}.  The partition function for thermal QCD
can be expressed in terms of a functional integral in  MQCD:
\begin{equation}
Z_{\rm QCD}(T) \;=\;
e^{- f_{\rm QCD}(\Lambda) \, T^3 \, V} \;
e^{- f_{\rm EQCD}(\Lambda,\Lambda') \, (g T)^3 \, V}
\int^{(\Lambda')} {\cal D} A_i^a
\exp \left( - \int d^3x {\cal L}_{\rm MQCD} \right).
\label{ZMQCD}
\end{equation}
In the second prefactor,
$f_{\rm EQCD} (gT)^3$  can be interpreted
as the coefficient of the unit operator in the effective lagrangian
for MQCD, which was omitted in (\ref{MQCD}).
Its dependence on  the ultraviolet cutoff $\Lambda'$ is cancelled by the
$\Lambda'$-dependence of the loop integrals in MQCD.

If we use dimensional regularization to cut off both the ultraviolet
and the infrared divergences in the perturbation expansion for EQCD,
then $-f_{\rm EQCD}(gT)^3$ is just the logarithm of the partition
function for EQCD.  It can be obtained to next-to-leading order in
$g$ by calculating 1-loop and  2-loop diagrams in EQCD.
Taking the $\overline{MS}$
scale of dimensional regularization to be $\Lambda$, the result is
\begin{equation}
f_{\rm EQCD} \; (gT)^3 \;=\;
{N_c^2 -1 \over 4 \pi} m_{\rm el}^3
\left\{ - {1 \over 3} \;+\; {N_c g_3 \over 16 \pi m_{\rm el}}
	\left[ {1 \over \epsilon}
	+ 4 \log {\Lambda \over 2 m_{\rm el}} + 3 \right] \right\}.
\label{fEQCD}
\end{equation}
Note that the pole in $\epsilon$ in (\ref{fEQCD}) cancels against
the pole in $f_{\rm QCD} T^3$, where $f_{\rm QCD}$ is given in (\ref{fQCD}).
Furthermore, after making the substitution (\ref{mel}) for $m_{\rm el}$
in (\ref{fEQCD}), the logarithms of $\Lambda$ also cancel.
The result is a logarithm of $T/m_{\rm el}$,
whose coefficient was first obtained
by Toimela \cite{toimela}.  In EQCD, the logarithm can be associated with
the renormalization of the operator $A_0^a A_0^a$, which mixes with
the unit operator.  Thus the coefficient $f_{\rm EQCD}$
of the unit operator in the effective lagrangian for EQCD
must have a dependence on $\Lambda$ that
cancels that of the renormalized operator $A_0^a A_0^a$.

In MQCD, perturbation expansions in the gauge coupling constant $g_3'$
are hopelessly plagued
with infrared divergences.  Thus the functional integral in (\ref{ZMQCD})
can only be calculated by nonperturbative methods, such as
lattice simulations of MQCD.  Surprisingly, $f_{\rm MQCD}$ can still
be expressed as a power series in $g(T)$. The most important momentum scale in
MQCD  is $g_3'$, which is of order $g^2T$.
The functional integral can therefore be expressed in the form
\begin{equation}
\int^{(\Lambda')} {\cal D} A_i^a
\exp \left( - \int d^3x {\cal L}_{\rm MQCD} \right)
\;\equiv\; e^{- f_{\rm MQCD}(\Lambda') \, (g^2 T)^3 \, V } ,
\label{IntMQCD}
\end{equation}
where $f_{\rm MQCD}$ is dimensionless. This function $f_{\rm MQCD}$
has a power series expansion in powers of $g(T)$.
The leading term of order $g^0$
comes from dropping the correction terms
$\delta {\cal L}_{\rm MQCD}$ in the effective lagrangian.  The resulting
functional integral is
\begin{equation}
\int {\cal D} A_i^a
\exp \left( - \int d^3x {1 \over 4} G_{ij}^a G_{ij}^a \right)
\;\equiv\; e^{- f^{(0)}_{\rm MQCD} \, (g_3')^6 \, V } .
\label{IntMQCD0}
\end{equation}
Since a pure nonabelian gauge theory in 3 space dimensions is an
ultraviolet finite theory, the only momentum scale is the gauge
coupling constant $g_3'$.  Thus, by dimensional analysis,
the coefficient $f_{\rm MQCD}^{(0)}$ is a pure number that can be
calculated using lattice simulations of the functional integral
(\ref{IntMQCD0}).  Higher order corrections to $f_{\rm MQCD}$
can be calculated by treating the terms in
$\delta {\cal L}_{\rm MQCD}$ as perturbations.  The leading
corrections comes from the dimension-5 operators
$g_3' f^{abc} G_{ij}^a G_{jk}^b G_{ki}^c$ and $(D_iG_{jk})^a (D_iG_{jk})^a$,
which have short-distance coefficients
proportional to $g_3^2/m_{\rm el}^3$.  The expectation values
of these operators are of order $(g_3')^{10}$, but they are
ultraviolet divergent and depend on the ultraviolet
cutoff $\Lambda'$ of MQCD.  This cutoff-dependence is cancelled by
the $\Lambda'$-dependence of $f_{\rm EQCD}$ and the parameters
of ${\cal L}_{\rm MQCD}$.  Thus the dimension-5
operators in $\delta {\cal L}_{\rm MQCD}$ give well-defined
corrections to $f_{\rm MQCD}$ that are of order $g^3$.

Combining (\ref{ZMQCD}) and (\ref{IntMQCD}), we obtain our final result
for the free energy density of a nonabelian gauge theory:
\begin{equation}
{F_{\rm QCD}(T)  \over T}
\;=\; f_{\rm QCD}(\Lambda) \; T^3
\;+\; f_{\rm EQCD}(\Lambda,\Lambda') \; (g T)^3
\;+\; f_{\rm MQCD}(\Lambda') \; (g^2 T)^3 ,
\label{F}
\end{equation}
where $g$ is the running coupling constant $g(T)$.
We have indicated the dependence of the functions on the
arbitrary factorization scales $\Lambda$ and $\Lambda'$,
which separate the momentum scales $T$, $gT$, and $g^2T$.
The coefficient $f_{\rm QCD}$ can be calculated using ordinary
perturbation theory in thermal QCD in the form of
a power series in $g^2(T)$, with coefficients
that are polynomials in $\log(T/\Lambda)$.
The coefficient $f_{\rm EQCD}$ can be calculated using
perturbation theory in EQCD,
which gives a power series in $g(T)$ with coefficients
that are polynomial in $\log(\Lambda/gT)$ and $\log(gT/\Lambda')$.
The coefficient $f_{\rm MQCD}$ can be expressed as a power series
in $g(T)$, with coefficients that can be calculated using
lattice simulations of MQCD.
The functions $f_{\rm QCD}$ and $f_{\rm EQCD}$ satisfy renormalization group
equations that can be used to sum up leading logarithms of $T/\Lambda$
and $gT/\Lambda'$, respectively.  By choosing $\Lambda$ of order $gT$ and
$\Lambda'$ of order $g^2T$, we can eliminate potentially large logarithms
of $1/g$ from the perturbative expansions of $f_{\rm EQCD}$ and $f_{\rm MQCD}$.

Adding (\ref{fQCD}) and (\ref{fEQCD}), with $g_3 = g \sqrt{T}$
and $m_{\rm el}$ given by (\ref{mel}), we recover the result for
the free energy of pure-glue QCD to order $g^4$ that was recently
obtained by Arnold and Zhai \cite{arnold-zhai}.  The order $g^5$
correction requires the calculation of $m_{\rm el}^2$ to next-to-leading
order in $g^2$ and the calculation of the 3-loop diagrams for the
partition function of EQCD.  A complete calculation to order $g^6$
requires calculating the 3-loop contribution to $f_{\rm QCD}$,
the 4-loop contribution to $f_{\rm EQCD}$, and the number
$f_{\rm MQCD}^{(0)}$ defined  by (\ref{IntMQCD0}).
This last number  can only be obtained with a
nonperturbative lattice calculation of the functional integral.

The problem of the perturbative infrared catastrophe of nonabelian
gauge theories is really one of providing a bridge between
perturbative calculations of the effects of the large momentum scale
$T$ and the nonperturbative calculations that are required
at the small momentum scale $g^2T$.  In this Letter,
we have provided such a bridge by constructing a sequence of two
effective field theories by first integrating out the momentum scale $T$
and then integrating out the momentum scale $gT$.
This method allows thermodynamic
quantities like the free energy to be calculated with arbitrarily
high accuracy in the high temperature limit.

\bigskip

This work was supported in part by the U.~S. Department of Energy,
Division of High Energy Physics, under Grant DE-FG02-91-ER40684.
I thank A. Nieto for valuable discussions.

%-------------------

\end{document}